\def\cm{cm$^{-1}$}
\def\bfa{Ba\-Fe$_2$As$_{2}$}

\def\bfca{Ba\-(Fe$_{0.92}$Co$_{0.08})_2$As$_{2}$}
\def\bfna{Ba\-(Fe$_{0.95}$Ni$_{0.05})_2$As$_{2}$}

\documentclass[aps,prl,showpacs,floatfix,preprint,superscriptaddress,byrevtex]{revtex4}
\usepackage{graphicx}
\usepackage{color}

\begin{document}
\title{Nodes in the Order Parameter of  Superconducting Iron Pnictides\\
Observed by Infrared Spectroscopy}
\author{D. Wu}
\author{N. Bari\v{s}i\'{c}}
\author{M. Dressel}
\affiliation{1.~Physikalisches Institut, Universit\"at Stuttgart,
Pfaffenwaldring 57, 70550 Stuttgart, Germany}
\author{G. H. Cao}
\author{Z. A. Xu}
\affiliation{Department of Physics, Zhejiang University, Hangzhou 310027, People's Republic of China}
\author{J. P. Carbotte}
\affiliation{Department of Physics and Astronomy, McMaster University, Hamilton, Ontario L8S 4M1, Canada}
\author{E. Schachinger}
\affiliation{Institut f\"ur Theoretische Physik, Technische Universit\"at Graz,
Petersgasse 16, 8010 Graz, Austria}
\date{\today}

\begin{abstract}
The temperature and frequency dependences of the conductivity are derived
from optical reflection and transmission measurements of electron doped BaFe$_2$As$_2$
crystals and films. The data is consistent with
gap nodes or possibly a very small gap in the crossover region between
these two possibilities. This can arise when one of the several pockets
known to exist in these systems has extended $s$-wave gap symmetry with
an anisotropic piece canceling or nearly so the isotropic part in some
momentum direction. Alternatively, a node can be lifted by impurity
scattering which reduces anisotropy. We find that the smaller gap on
the hole pocket at the $\Gamma$ point in the Brillouin zone is isotropic
$s$-wave while the electron pocket at the $M$ point has a larger gap
which is anisotropic and falls in the crossover region.
\end{abstract}

\maketitle
It took almost ten years from the discovery of
high-temperature superconductivity in cuprates to the
confirmation of $d$-wave symmetry of the order parameter and its
general acceptance. While only phase sensitive tunneling
experiments could prove the change in sign of the wave function
\cite{Tsui94}, previous experiments on the electrodynamic
properties provided compelling evidence for nodes in the gap
\cite{Hardy93}. When the new iron-based superconductors
were discovered in 2008 \cite{Kamihara08}, the momentum dependence
of the gap function $\Delta({\bf k})$ was raised as one of the
most important questions of fundamental relevance
\cite{Kivelson08,Mazin10}. The magnitude of the gap and the
quantum mechanical phase of the electron pairs in ${\bf k}$ space
provides information needed to identify the
pairing mechanism. Compared to cuprates,
the situation here is more
complicated since five bands are close to this Fermi level
from which two are most relevant for the physical behavior
\cite{Fink09,Mazin08}.

Various experimental methods have been applied to elucidate the
gap structure of iron pnictides (for recent reviews, see Refs.
\cite{Ishida09,Mazin10,Johnston10}); however, the results are often
contradictory even within the doped \bfa\ compounds of the 122
family: Specific heat \cite{Gofryk10,Mu10}, heat transport
measurements \cite{Dong10,Tanatar10}, the temperature dependence
of the penetration depth \cite{Hashimoto09,Martin09,Mishra09},
tunneling spectroscopy (STM) \cite{Szabo09,Samuely09}, spin-lattice
relaxation \cite{Fukazawa09,Nakai09,Zhang10},  muon spin rotation
($\mu$SR) \cite{Hiraishi09,Goko09,Khasanov09} and electronic Raman
scattering \cite{Muschler09} are discussed controversially, since
the results might depend on the sample quality, doping regime or
particular system. Angle-resolved photoemission spectroscopy
(ARPES) consistently detects nodeless, isotropic superconducting
gaps on all sheets of the Fermi surface \cite{Terashima09,Evtushinsky09}
but might be hampered by resolution problems. According to their
interpretation, gaps open simultaneously below $T_c$ with two distinct
energies on different parts of the Fermi surface: approximately 5~meV
($\Gamma$ point) and 12~meV ($\Gamma$ point and M point).
There is also a second pocket centered at $\Gamma$ with a smaller
radius and a gap of $12\,$meV but often this is not included in
the simplest two pocket model as we will employ here.
This is taken as evidence for an $s$-wave pairing state
\cite{Mazin08,Mazin10} but remains silent as to whether or not the state
is of $s^{\pm}$-type, i.e., the
symmetry of the order parameter reverses sign for electron and
hole pockets on the Fermi surface. On the other
hand new phase sensitive scanning tunneling microscopy (STM) data
\cite{hanaguri10} have provided strong evidence for $s^\pm$ symmetry
which we will assume to be the case in this paper.

Optical experiments might contribute to clarify these issues for
the superconducting condensate and electronic excitations can be
probed with very high resolution
over a wide energy range. Hence, microwave, THz and infrared methods have
been the primary tool to investigate the energy gap of
conventional superconductors \cite{Tinkham96,DresselGruner02}, and
have been applied to iron pnictides, in particular to 122
compounds by single crystal reflection
\cite{Li08SC,Hu09c,Wu10b,Wu10c,Kim09,Heumen09,Barisic10} and films transmission
\cite{Gorshunov10,Perucchi10,Fischer10} experiments.

We have measured the optical reflectivity of \bfca\ and \bfna\
single crystals ($T_c=25$ and 20~K) off the $ab$ surface
over a wide frequency and
temperature range  \cite{Wu10b,remark3}. The complex
conductivity $\hat{\sigma}=\sigma_1 +{\rm i}\sigma_2$ is calculated
via Kramers-Kronig analysis and further analyzed by the extended
Drude model \cite{DresselGruner02} in order to obtain the frequency dependent optical
scattering rate $\tau^{-1}(\omega)$ related to the total optical
conductivity by $\tau^{-1}(\omega) = {\omega_p}^2{\rm Re}\{\hat{\sigma}^{-1}(\omega)\}/
(4\pi)$ with $\omega_p$ the plasma frequency.

\begin{figure}
 \centering
 \includegraphics[width=0.45\columnwidth]{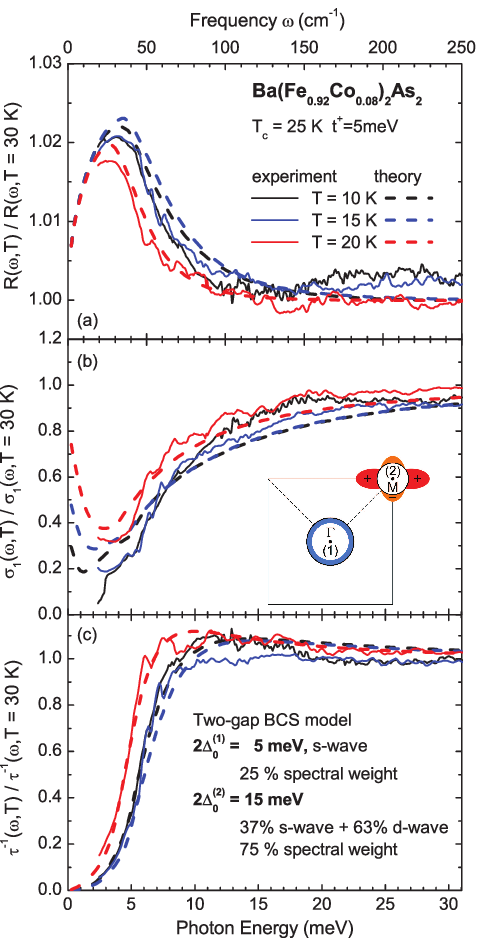}
 \caption{\label{fig:Fig1} (Color online) Relative change of the optical properties of \bfca\ upon entering the
superconducting state at $T_c=25$~K. The solid lines correspond to
experimental data obtained at $T=10$, 15 and 30~K. The dashed
lines represent fits by a two gap model with different symmetry,
where the first gap $2\Delta_0^{(1)}=5$~meV has a simple $s$-wave
symmetry, while the second gap exhibits $s+d$ wave symmetry with
37\%\ $s$-wave character  and an amplitude
$2\Delta_0^{(2)}=15$~meV.
 (a)~Reflectivity $R(\omega,T)/R(\omega,T=30\,{\rm K})$,
 (b)~optical conductivity $\sigma_1(\omega,T)/\sigma_1(\omega,T=30\,{\rm K})$,
 and
 (c)~frequency dependent optical scattering rate
 $\tau^{-1}(\omega,T)/\tau^{-1}(\omega,T=30\,\textrm{K})$. The inset sketches
 the two Fermi-surface pockets in {\bf k} space, with an isotropic $s$-wave
 gap about $\Gamma$ and an anisotropic extended $s$-wave gap about $M$.}
\end{figure}

In Fig.~\ref{fig:Fig1}(a) the change in the optical reflectivity
of \bfca\ is plotted for different temperatures below $T_c=25$~K.
The maximum in ${R(\omega,T)}/{R(\omega,T=30~{\rm K})}$ at
approximately 5~meV  is a clear indication that a
superconducting gap of this energy opens. However, looking at
the corresponding conductivity plotted in panel (b), it becomes
obvious that $\sigma_1(\omega,T)/\sigma_1(\omega,T=30~{\rm K})$
does not drop to zero at the gap as rapidly as expected for a
simple $s$-wave superconductor \cite{Tinkham96,DresselGruner02},
but decreases in an almost linear fashion before it increases
again at very low energies. This is the hallmark of nodes in the
gap for which excitations are possible at infinitely small
energies. The corresponding optical scattering rate is plotted
Fig.~\ref{fig:Fig1}(c).

In order to simultaneously describe these optical properties we
decompose the conductivity in two contributions
$\hat{\sigma}=\hat{\sigma}^{(1)}+\hat{\sigma}^{(2)}$, corresponding to
a hole band [$\omega_{p}^{(1)}/(2\pi c)\approx$ 0.58~eV] and an
electron band [$\omega_{p}^{(2)}/(2\pi c)\approx$ 1~eV].
In order to keep the number of fit parameters minimal
we have chosen a single scattering rate for both terms \cite{remark1}.
The superconducting state is mimicked by a two-gap
BCS model with an isotropic $s$-wave gap $2\Delta_0^{(1)}=5$~meV
on the hole pocket, while the gap on the electron
pocket was modeled with an extended $s$-wave gap which when referred
to the $M$ point as origin takes on the form
$\Delta^{(2)}=\Delta_s+\Delta_d\sqrt{2}\cos\{2\theta\}$
($\Delta_s$ is the $s$-wave component, $\Delta_d$ is the amplitude
of the $d$-wave component, and $\theta$ is the polar angle on the
Fermi surface). Here $\Delta_s=\alpha\Delta_0^{(2)}$ and
$\Delta_d=\sqrt{1-\alpha^2}\Delta_0^{(2)}$, where
$\Delta_0^{(2)}=\sqrt{\langle[\Delta^{(2)}(\theta)]^2\rangle_{\theta}}$,
$\langle\ldots\rangle_{\theta}$ the Fermi-surface average and
$2\Delta_0^{(2)}=15$~meV. The percentage of $s$-wave component is
expressed by $x=\alpha/(\alpha+\sqrt{1-\alpha^2})$
\cite{Schachinger09}. We refer to this model as extended $s^\pm$.

As seen from Fig.~\ref{fig:Fig1}, the fit to the experimental data
and derived quantities is remarkably good; even small details and
particular features are described well, such as the increase in
$R(\omega,T)/R(\omega,30~{\rm K})$ as the temperature is lowered,
the shift of the peak to higher frequencies, the gradual drop in
$\sigma_1(\omega,T)/\sigma_1(\omega,30~{\rm K})$ without
approaching zero, but showing an upturn as $\omega\rightarrow 0$,
the slight increase of the scattering rate as $\omega$ decreases
with a rapid drop when the smaller gap is reached, but also the way
it approaches zero. The consistent description of all quantities
in the complete frequency and temperature range with a single set
of parameters gives confidence in the proper choice of the
model. In all calculations the temperature dependence of the
superconducting gaps was modeled by the BCS mean field
dependence.

The comparison with optical data obtained by transmission
experiments of thin films of Co doped \bfa\
\cite{Gorshunov10,Fischer10} corroborates our hypothesis. The
pronounced coherence peak observed in the microwave conductivity
$\sigma_1(T)$ below $T_c$ is
significantly broader than expected from simple $s$-wave BCS
theory \cite{Tinkham96,DresselGruner02} and extends to much lower
temperatures; but it can be perfectly reproduced by the present
model with basically the same set of parameters \cite{remark2}.
This also includes the quadratic temperature   dependence of the
penetration depth $\lambda\propto T^2$ at the lowest temperatures
(see also Modre {\it et al.} \cite{modre98}).

\begin{figure}[tp]
 \includegraphics[width=0.5\columnwidth]{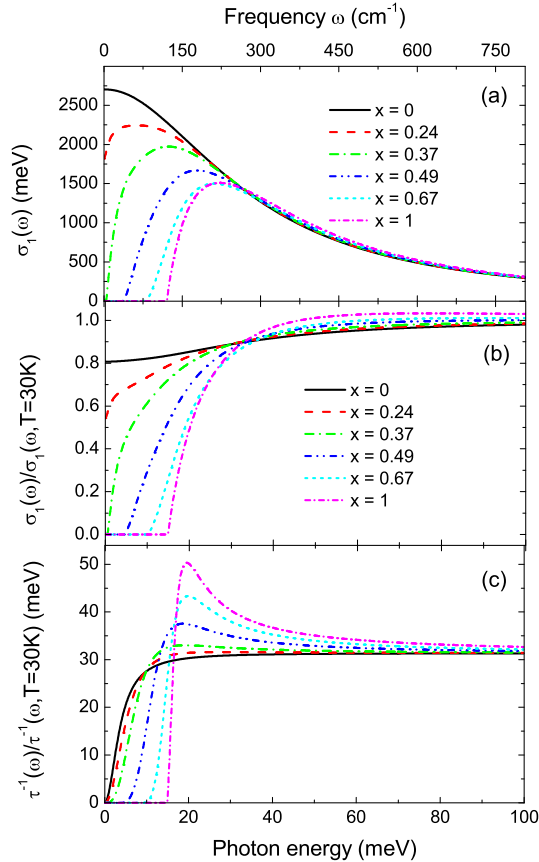}
  \caption{(Color online) (a) The real part of the optical conductivity
$\sigma_1(\omega)$ in meV vs the photon energy $\omega$ in meV for
an $s+d$-wave model at $T=1.25\,$K and for various values of the
parameter $x$ which multiplied by hundred gives the percentage of
the $s$-wave admixture. Thus, $x=0$ corresponds to pure $d$-wave
symmetry of the gap while $x=1$ corresponds to $s$-wave symmetry.
$\Delta_0 = 7.5\,$meV and the elastic scattering rate
$\tau^{-1}_{\rm imp} = 31.4\,$meV. The $x=0.37$ corresponds to the
$s$-wave admixture used in the analysis presented in Fig.~\ref{fig:Fig1}.
(b) The same as (a) but for the normalized real part of the optical
conductivity $\sigma_1(\omega)/\sigma_1(\omega, T=30\,\textrm{K})$ with
$\sigma_1(\omega, T=30\,\textrm{K})$ its normal state value. (c) The same
as (a) but for the normalized optical scattering rate
$\tau^{-1}(\omega)/\tau^{-1}(\omega,T=30\,\textrm{K})$.
}
  \label{fig:Fig2}
\end{figure}

To better understand our results it is important to consider very
low temperatures considerably below the lowest experimental temperature
given in Fig.~\ref{fig:Fig1}. In Fig.~\ref{fig:Fig2}
we show results for an extended $s$-wave gap with the same
parameters as we have used in our experimental fit \cite{remark1}
with the reduced temperature set at $t=T/T_c=0.05$, i.e. $T=1.25\,$K.
Fig.~\ref{fig:Fig2}(a) shows $\sigma_1(\omega)$ in meV vs $\omega$ also
in meV. The various curves are for six different values of
the variable $x$ which multiplied by hundred gives the percentage of
the $s$-wave component in an $s+d$-wave admixture as labeled.
The case $x=0.37$ corresponds to the anisotropy we used in our fit
to experiment. We see that this corresponds to a very small but finite
spectral gap in the real part of the optical conductivity. This gap
is so small, however, that it falls near the end of the available THz
range and is not present at $10\,$K where we have a gapless behavior. In
Fig.~\ref{fig:Fig2}(b) we show the same results but normalized to the
real part of the optical conductivity $\sigma_1(\omega,T=30\,\textrm{K})$
at $T=30\,$K in the normal state while in Fig.~\ref{fig:Fig2}(c) we show
$\tau^{-1}(\omega)/\tau^{-1}(\omega,T=30\,\textrm{K})$. It is clear from
these results that it would be very difficult even at low temperatures
to differentiate between nodes and the case of a small spectral
gap. All we can firmly conclude from our present analysis is that the
material studied falls close to the crossover region between these
two possibilities.
\begin{figure}
 \centering
 \includegraphics[width=0.45\columnwidth]{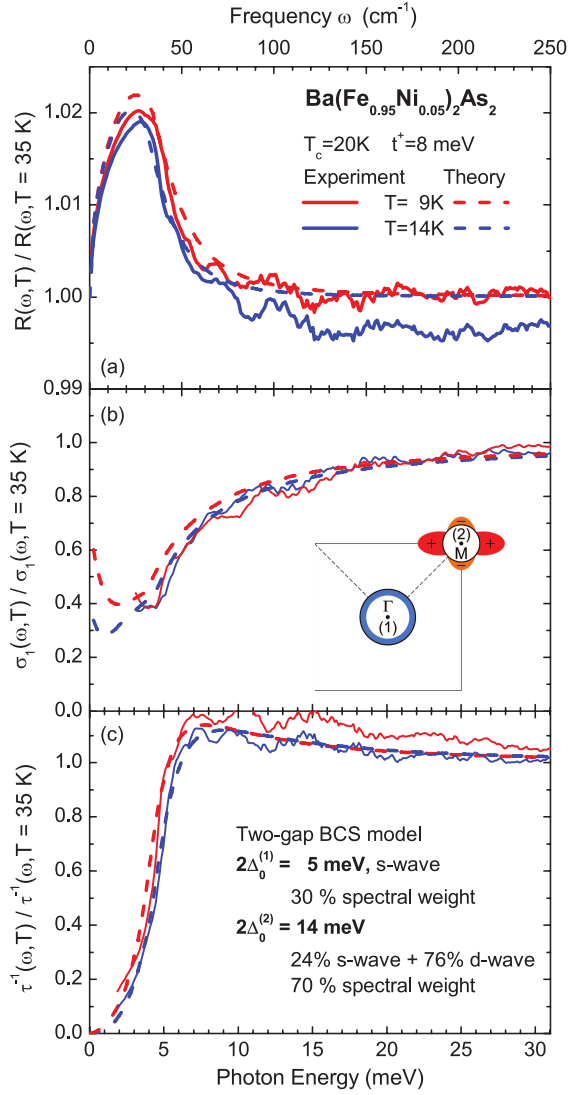}
 \caption{\label{fig:Fig3} (Color online) Relative change of the optical properties of \bfna\ upon entering the
superconducting state at $T_c=20$~K. The solid lines correspond to
experimental data obtained at $T=9$ and 14~K. The dashed lines
represent fits by a two gap model with different symmetry, where
the first gap $2\Delta_0^{(1)}=4.5$~meV has a simple $s$-wave
symmetry, while the second gap exhibits $s+d$ wave symmetry with
24\% $s$-wave character and an
amplitude $2\Delta_0^{(2)}=14$~meV.
 (a)~Reflectivity $R(\omega,T)/R(\omega,T=35\,{\rm K})$,
 (b)~optical conductivity $\sigma_1(\omega,T)/\sigma_1(\omega,T=35\,{\rm K})$,
 and (c)~frequency dependent optical scattering rate
$\tau^{-1}(\omega,T)/\tau^{-1}(\omega,T=35\,{\rm K})$. }
\end{figure}

In order to proof the general applicability of our description, we
have analyzed the optical data for \bfna\ in a similar fashion.
Fig.~\ref{fig:Fig3} displays the corresponding results for $T=9$
and 14~K, referenced to the normal state data measured at
$T=35$~K. In this case the extended $s^\pm$ model is
defined by the isotropic $s$-wave gap of $2\Delta_0^{(1)}=4.5$~meV
for 30\%\ of the spectral weight, while the dominant electron band is gapped
of size $2\Delta_0^{(2)}=14$~meV with $x = 0.24$
and contains nodes.The distribution of spectral weight
between both gaps is $w = w^{(1)}+w^{(2)} = 0.3 + 0.7$.
The plasma frequency was set to $\omega_p = 1.3$~eV and
the elastic scattering rate to $\tau^{-1}_{\rm imp} = 50$~meV.
Again, the agreement between experiment and theory is very good, considering
that the lower energy scale approaches the limit of the frequency
range accessible by experiment.

In conclusion, we have analyzed the temperature and frequency
dependences of the optical properties of electron doped BaFe$_2$As$_2$ in
the superconducting state. From the low-frequency reflectivity and
conductivity ratios $R_s/R_n$ and $\sigma_s/\sigma_n$ below and
above $T_c$, we conclude the symmetry of the order parameter
can be described by an extended $s^\pm$ model.
Two gaps are needed to get agreement. The smaller gap is
isotropic $s$-wave while the larger gap is anisotropic with nodes,
or possibly with a very small spectral gap. When referred to the
$M$ point in the Brillouin zone as origin it is described as a mixture
of $s$ and $d$-wave with the $d$-wave component providing a mechanism
for gap nodes or a very small gap in certain directions in momentum space.
This is strongly  supported by a
broad coherence peak in the microwave conductivity $\sigma_1(T)$
and the temperature dependence of the penetration depth
$\lambda(T)\propto T^2$ observed in optical transmission measurements
through films.
Our theoretical two gap model gives $\lambda(T)\propto T^{1.9}$
which is in very good agreement with experiment.
Finally, we note that optics cannot distinguish
between gap phases (it is not phase sensitive) but recent STM
experiments \cite{hanaguri10} firmly established the $s^\pm$ symmetry
with electron and hole pockets having the opposite sign.

We appreciate discussions with D.N. Basov, N. Drichko, C. Haule, D. van der Marel and A.V. Pronin.
N.B. and D.W. acknowledge a fellowship of the Alexander von Humboldt-Foundation. G. H. Cao and Z-A. Xu acknowledge partial support from NSFC.


\begin{thebibliography}{99}
\bibitem{remark3} We measured the optical conductivity down to 20 \cm\ using IR 1.5K Bolometer. The Hagen-Rubens extrapolation was applied for the low frequency. Below 35 \cm\, we used a proper smoothing procedure to reduce the experimental noise.
\bibitem{remark1}We have chosen $\epsilon_{\infty}=10$ to account
for higher-energy excitations.
The total plasma frequency is $\omega_p/(2\pi c)=1.15$~eV
and the elastic scattering rate $\tau_{\rm imp}^{-1}= 31.4$~meV.
The distribution of spectral weight between both components is
$w=w^{(1)}+w^{(2)}=0.25+0.75=1$, and $x=0.37$.
\bibitem{remark2} The energy gaps for the Ba(Fe$_{0.9}$Co$_{0.1}$)$_2$As$_2$
film ($T_c\approx 25\,$K) are $2\Delta_0^{(1)}= 6\,$meV and
$2\Delta_0^{(2)}= 16\,$meV. The distribution of spectral weight is the same as found for the crystal: $w^{(1)}=0.25$ and $w^{(2)}=0.75$ with
$\alpha=0.45$ corresponding to an $s$-wave component of  $x=1/3$. See details in Ref.[40]
\end{thebibliography}
\end{document}